\documentclass[12pt]{article}

\begin{document}

\title{ Lorentz violation, Two times physics and Strings}

\author{Juan M. Romero\thanks{jromero@correo.cua.uam.mx}
\\[0.5cm]
\it Departamento de Matem\'aticas Aplicadas y Sistemas,\\
\it Universidad Aut\'onoma Metropolitana-Cuajimalpa\\
\it M\'exico, D.F  01120, M\'exico\\[0.3cm] \\
 Oscar S\'anchez-Santos
\thanks{oscarsanbuzz@yahoo.com.mx} and Jos\'e David Vergara
\thanks{vergara@nucleares.unam.mx}
\\[0.5cm]
\it Instituto de Ciencias Nucleares,\\
\it Universidad Nacional Aut\'onoma de M\'exico,\\
\it Apartado Postal 70-543, M\'exico 04510 DF, M\'exico
} 

\date{}

\pagestyle{plain}

\maketitle

\begin{abstract}
We study a recently proposed generalization of the relativistic
particle  by Kosteleck\'y, that includes explicit Lorentz violation.
We present an alternative action for this system and we show that
this action can be interpreted as a particle in curved space with a
metric that depends on the Lagrange multipliers. Furthermore, the
following results are established for this model: (i) there exists a
limit where this system has more local symmetries that the usual
relativistic particle; (ii) in this limit if we restore the Lorentz
symmetry we obtain a direct relationship with the two time physics;
(iii) also we show that if we intent  to restore the Poincar\'e
symmetry we obtain the action of the relativistic bosonic string.

\end{abstract}

\section{Introduction}
\label{s:Intro}
In several recent publications have been proposed different systems
with explicit Lorentz violation. Since, the Lorentz symmetry is one
of the cornerstones of the Laws of Physics, these works have
attracted a lot of attention. For example, the Ho\v{r}ava's gravity
breaks locally this symmetry, but has as consequence a
renormalizable gravity \cite{horava:gnus}. In Quantum Field Theory
there are also several studies that consider this symmetry violation
\cite{kostelecky:gnus}, a review on this subject can be found in
\cite{kostelecky0:gnus}. An interesting point is that various models
that break the Lorentz symmetry, at the level of Quantum Field
Theory \cite{lehnert:gnus}, imply a dispersion relation of the form
\begin{eqnarray}
\left(P_{\mu}+a_{\mu}\right)^{2}+m^{2}+b\cdot b \mp 2\sqrt{ \left(P\cdot b+a\cdot b\right)^{2}+m^{2}b\cdot b}=0,
\label{eq:RD}
\end{eqnarray}
with $a$ and $b$ two constant vectors. Clearly this dispersion
relation violates the Lorentz symmetry. Recently in Ref.
\cite{kostelecky1:gnus} was found a mechanical model that implies
the dispersion relation (\ref{eq:RD}). This model  allows to
understand more deeply the kind of phenomena that are involved in
the Lorentz  symmetry violation. An interesting result is the
relationship between this model and the Finsler geometry
\cite{kostelecky2:gnus}. Let us also mention here that recently the
Finsler geometry was proposed as an alternative to the Minkowski
geometry in the High Energy Physics regime close to the Planck scale
\cite{finsler:gnus}. In that sense, independently of the origin of
the proposed action of \cite{kostelecky1:gnus}, this system has very
interesting properties that are worth to study.\\

The purpose of this work is to study several aspects of the action
presented in \cite{kostelecky1:gnus}, and show that this system has
several interesting properties. In particular, we introduce an
alternative action to the proposed by Kosteleck\'y, et al
\cite{kostelecky1:gnus}. Furthermore, we establish that this action
is equivalent to a particle in a curved space, where the metric
depends on two Lagrange multipliers. This result exhibit that
Finsler geometry could be studied in this way. We also consider the
local symmetries of the system and we show that there exist a limit
where the model has more local symmetries that the classical
relativistic free particle. As another property, we show that if in
this limit we tray to recover the Lorentz symmetry  we obtain
several generalized Lorentz invariant systems. These systems have
interesting local symmetries. For example, we show in Sec. 5 that
applying the Dirac's method \cite{Dirac:gnus}, we obtain the action
for the two time physics. One of the hallmarks of the two time
physics is that in only one action are unified different models to
the level of particle. Consequently this theory acts as a
unification model to the level of point particle \cite{bars:gnus}.
Also, we show that, nevertheless the Lorentz symmetry is recovered
we still don't have a
restoration of the Poincar\'e symmetry.\\

Now, the dispersion relation (\ref{eq:RD}) originally appears from a
Field Theory, and the action for the point particle is obtained from
a simplification of a given Field Theory. It is worth to notice that
the same happens in the case of the action for the two time physics,
where it is obtained from a reduction of a Field Theory
\cite{marnelius:gnus}. In this work to recover the Poincar\'e
invariance, we will take the inverse path, i.e. we will transform
our point particle action into a field theory. We will see that if
we consider that the coordinates depend on other parameter
  $\sigma,$ i.e.  $X^{M}= X^{M}(\tau,\sigma),$ and instead of
 $b$ we take $T \frac{ \partial X^{M}}{\partial \sigma}$ the system will be invariant
 under  Lorentz and Poincar\'e transformations. We also show that in
 this case the action of the relativistic bosonic string is
 obtained. We must mention that in the case of the Snyder space \cite{snyder:gnus},
 Yang proposed an extra dimension in such way to become the Snyder space
 Poincar\'e invariant \cite{yang:gnus}. A similar process will be
 taken in this work to recover the Lorentz and Poincar\'e
 symmetries.\\

This work is organized in the following way: In Section 2 we perform
a canonical analysis of the action proposed by Kosteleck\'y
\cite{kostelecky2:gnus}. In section 3 we find an alternative action
to such system. In Section 4 we study the case where the
perturbation is stronger that the usual term. In Section 5 we
recover the Lorentz symmetry and we establish a relationship between
this system and the two time physics. For the Section 6, we show
that in order to recover the Poincar\'e symmetry we obtain the
relativistic string. Finally we summarize our results in Section 7.

\section{Action for the system}

The action proposed in Ref. \cite{kostelecky1:gnus} is
\begin{eqnarray}
 S=\int d\tau \left( -m\sqrt{- \dot X\cdot \dot X}- a\cdot \dot X \pm
 \sqrt{ \left(b \cdot \dot X\right)^{2}-\left(b\cdot b\right) \left(\dot X \cdot \dot X \right)}\right).
\label{eq:accion}
\end{eqnarray}
where $A\cdot A=A_{M}A^{M}=\eta_{MN}A^{M}A^{N}$ with $N,
M=0,1,2,\cdots, D$ and ${\rm sig} (\eta_{MN})=(-1,1,\cdots, 1).$
Furthermore $a_{M}$ and $b_{M}$ are constant vectors.  Note that
this action is invariant under reparametrizations given by
$\frac{dX}{d\tau}=\frac{d \tilde \tau }{d\tau} \frac{dX}{d\tilde \tau}.$ \\

The canonical momenta of the system are
\begin{eqnarray}
P_{M}=m\frac{\dot X_{M} }{\sqrt{-\dot X  \cdot \dot X}}-a_{M} \pm
 \frac{  \left( b \cdot \dot X\right) b_{M} -\left(b \cdot b\right) \dot X_{M} }
 { \sqrt{ \left(b \cdot \dot X\right)^{2}-\left(b \cdot b\right) \left(\dot X \cdot \dot X\right)}},
\end{eqnarray}
from these we obtain the canonical Hamiltonian
\begin{eqnarray}
H_{c}=P \cdot \dot X-L=0.
\end{eqnarray}
Also, we see that the following relations are satisfied
\begin{eqnarray}
\left(P +a\right)\cdot b &=& m\frac{\dot X \cdot  b }{\sqrt{-\dot X \cdot \dot X}}, \label{eq:1}\\
\left(P_{M}+a_{M}\right)^{2}&=&-m^{2}\pm \frac{2m\sqrt{\left(b \cdot \dot X\right)^{2}-
\left(b \cdot b\right) \left(\dot X \cdot  \dot X\right) }}{\sqrt{-\dot X \cdot \dot X}} -b\cdot b, \label{eq:2}
\end{eqnarray}
introducing (\ref{eq:1}) in  (\ref{eq:2}), we found
\begin{eqnarray}\label{disp1}
\left(P_{M}+a_{M}\right)^{2}+m^{2}+b\cdot b \mp 2\sqrt{ \left(P\cdot
b+a\cdot b\right)^{2}+m^{2}b\cdot b}=0.
\end{eqnarray}
The dispersion relation (\ref{disp1}) is consistent with several
models developed to test a possible violation of the Lorentz
symmetry \cite{lehnert:gnus}.\\

Using the Dirac's method \cite{Dirac:gnus,henneaux:gnus}, the total
Hamiltonian is
\begin{eqnarray}
H_{T}&=&\lambda \Phi, \\
 \Phi&=&\frac{1}{2}\left[\left(P_{M}+a_{M}\right)^{2}+m^{2}+b\cdot b \mp 2
 \sqrt{ \left(P\cdot b+a\cdot b\right)^{2}+m^{2}b\cdot b}\right],
\end{eqnarray}
where $\lambda$ is Lagrange multiplier. The Hamiltonian action
results
\begin{eqnarray}
S=\int d\tau  \left( P \cdot \dot X -\lambda \Phi\right).
\end{eqnarray}

In the next section we will show that exist an alternative action
for this system.

\section{Action without square roots}

For the relativistic particle and inclusive for the string and
membranes a more suitable action for describing their variational
principles is to avoid the introduction of the square root by
including a Lagrange multiplier. In the case of the point-particle
with Lorentz-violation (\ref{eq:accion}) we have two independent
square roots so we need to introduce a pair of Lagrange multipliers.
The resulting action is given by

\begin{equation}\label{Altact}
S = \int {d\tau \left\{ {\frac{1}{2}\left[ {\frac{{{{\dot
X}^2}}}{\lambda } - \lambda {m^2} - 2a \cdot \dot X \pm
\frac{{{{\left( {b \cdot \dot X} \right)}^2} - {b^2}\left( {\dot X
\cdot \dot X} \right)}}{\beta } \pm \beta } \right]} \right\}}.
\end{equation}
This action is equivalent to (\ref{eq:accion}) if we use the
equations of motion of the two Lagrange multipliers.  The above
action (\ref{Altact}) can be rewritten as
\begin{equation}\label{Altact1}
S = \int {d\tau } \left\{ {\frac{1}{2}\left[ {{g_{MN}}{{\dot
X}^M}{{\dot X}^N} - 2a \cdot \dot X - \lambda {m^2} \pm \beta }
\right]} \right\},
\end{equation}
where the metric $g_{MN}$ is a deformation of the standard Minkowski
metric given by
\begin{equation}\label{metri}
{g_{MN}} = \left( {\frac{{\beta  \mp \lambda {b^2}}}{{\lambda \beta
}}} \right){\eta _{MN}} \pm \frac{{{b_M}{b_N}}}{\beta }.
\end{equation}
Using the action (\ref{Altact1}) is easy to see that in the limit
$b\to 0$ we recover the relativistic particle coupled to external
electromagnetic field $a.$ To compute the Hamiltonian we use the action
(\ref{Altact1}) and we get the momenta
\begin{equation}\label{mome}
{p_M} = {g_{MN}}{\dot X^N} - {a_M}.
\end{equation}
Then the canonical Hamiltonian is
\begin{equation}\label{HC}
{H_C} = \frac{{{g^{MN}}}}{2}\left( {{P_M} + {a_M}} \right)\left(
{{P_N} + {a_N}} \right) + \frac{1}{2}\left( {\lambda {m^2} \mp \beta
} \right),
\end{equation}
with the inverse metric $g^{MN}$ results
\begin{equation}\label{Invmetri}
{g^{MN}} = \frac{{\lambda \beta }}{{\beta  \mp {b^2}\lambda }}{\eta
^{MN}} \mp \frac{{{\lambda ^2}}}{{\beta  \mp \lambda
{b^2}}}{b^M}{b^N}.
\end{equation}
Furthermore we have two primary constraints
\begin{equation}\label{twoc}
{p_\lambda } \approx 0,\quad \quad {p_\beta } \approx 0,
\end{equation}
and in consequence the total Hamiltonian is
\begin{equation}\label{HT}
{H_T} = {H_C} + {\mu _1}{p_\lambda } + {\mu _2}{p_\beta }.
\end{equation}
From the evolution of the primary constraints we get

\[\begin{array}{l}
{{\dot p}_\lambda } = \left\{ {{p_{\lambda ,}}{H_T}} \right\} =
 - \frac{{\partial {g^{MN}}}}{{\partial \lambda }}\frac{1}{2}\left( {{P_M} + {a_M}}
  \right)\left( {{P_N} + {a_N}} \right) - \frac{{{m^2}}}{2},\\
{{\dot p}_\beta } = \left\{ {{p_{\beta ,}}{H_T}} \right\} =  -
\frac{{\partial {g^{MN}}}}{{\partial \beta }}\frac{1}{2}\left(
{{P_M} + {a_M}} \right)\left( {{P_N} + {a_N}} \right) \pm
\frac{1}{2}.
\end{array}\]
That results in the two following conditions
\begin{equation}\label{c1}
\left( {{\beta ^2}{\eta ^{MN}} + \left( {{b^2}{\lambda ^2} \mp
2\lambda \beta } \right){b^M}{b^N}} \right)\left( {{P_M} + {a_M}}
\right)\left( {{P_N} + {a_N}} \right) +
 {m^2}{\left( {\beta  \mp {b^2}\lambda } \right)^2} \approx 0,
\end{equation}
\begin{equation}\label{c2}
\left( { \pm {b^2}{\eta ^{MN}} \mp {b^M}{b^N}} \right)\left( {{P_M}
+ {a_M}} \right)\left( {{P_N} + {a_N}} \right) \pm {\left( {\beta
\mp {b^2}\lambda } \right)^2} \approx 0.
\end{equation}
Solving from (\ref{c2}) for  $\beta$ we get
\begin{equation}\label{beta}
\beta  = \lambda \left( { \pm {b^2} + \sqrt {A} } \right),
\end{equation}
with $A$ given by
\begin{equation}\label{A}
A = \left( { - {\eta ^{MN}}{b^2} + {b^M}{b^N}} \right)\left( {{P_M}
+ {a_M}} \right)\left( {{P_N} + {a_N}} \right).
\end{equation}
It should be noted that we can rewrite the metric in terms of the
momenta or the velocities and it is interesting to observe that at
this moment it is not of the Finsler type, since we still haven't
eliminated the Lagrange multiplier $\lambda$. For the metric we
obtain,
 ¿\begin{equation}\label{metri1}
{g_{MN}} = \frac{1}{{\lambda \left( { \pm {b^2} + \sqrt A }
\right)}}\left( {\sqrt A {\eta _{MN}} \pm {b_M}{b_N}} \right).
 \end{equation}
 Furthermore, from (\ref{c1}) using $\beta$ given by (\ref{beta}) we recover the constraint given in (6).

\section{Strong perturbation limit}

The action (\ref{eq:accion}) has been proposed to study the Lorentz
symmetry breaking, where the usual term is greater than the
perturbation that breaks the symmetry. But it is interesting to
study what happen at the contrary, i.e., if the correction is bigger
than the usual relativistic term. In other words we are considering
the ultrarelativistic regime, where we assume that $a_{M}, b_{M}$
are such that
\begin{eqnarray}
\left |m\sqrt{- \dot X\cdot \dot X}\right| << \left|-a\cdot \dot X \pm
 \sqrt{ \left(b \cdot \dot X\right)^{2}-\left(b\cdot b\right) \left(\dot X \cdot \dot X \right)}\right|,
\end{eqnarray}
in this regime we obtain the action
\begin{eqnarray}
 S=\int d\tau \left( - a\cdot \dot X \pm
 \sqrt{ \left(b \cdot \dot X\right)^{2}-\left(b\cdot b\right) \left(\dot X \cdot \dot X \right)}\right).
 \label{eq:accion1}
\end{eqnarray}
For the canonical momenta we have
\begin{eqnarray}
P_{M}=a_{M} \pm \frac{  \left(b\cdot \dot X\right) b_{M} -\left(b
\cdot b\right) \dot X_{M} } { \sqrt{ \left(b\cdot \dot
X\right)^{2}-\left(b \cdot b\right) \left(\dot X \cdot\dot
X\right)}},
\end{eqnarray}
and now these satisfy two primary constraints
\begin{eqnarray}
\Phi_{1}&=&\left(P+a\right)\cdot b=0,\\
\Phi_{2}&=&\left(P_{M}+a_{M}\right)^{2}+b\cdot b=0.
\end{eqnarray}
In this way, the total Hamiltonian is
\begin{eqnarray}
H=\lambda_{1}\Phi_{1}+ \lambda_{2}\Phi_{2}.
\end{eqnarray}
The constraints $ \Phi_{1}$  and $\Phi_{2}$ are first class, since
\begin{eqnarray}
\{\phi_{1},\phi_{2}\}=0.
\end{eqnarray}
This shows that the action (\ref{eq:accion1}) has more local
symmetries than the original action (\ref{eq:accion}). In this case
the gauge symmetries are given by
\begin{eqnarray}
\delta_{1} X_{M}&=& \epsilon_{1}(\tau) b_{M},\quad \delta P_{M}=0,\quad \delta \lambda_{1}=
\dot \epsilon_{1}(\tau),\nonumber\\
\delta_{2} X_{M}&=& \epsilon_{2}(\tau) 2\left(P_{M}+a_{M}\right),\quad \delta P_{M}=0,\quad
\delta \lambda_{2}=\dot \epsilon_{2}(\tau).
\end{eqnarray}

\section{Lorentz symmetry and two time physics}
By eliminate the usual term $m\sqrt{- \dot X\cdot \dot X}$ in the
action (\ref{eq:accion}) we have lost the usual relativistic
particle. However, we have more local symmetries. Now, we will see
that by recovering the Lorentz symmetry we will get still more local
symmetries and we can related this system to the action of two time
physics. \\

By simplicity we consider  $a_{M}=0,$ in this case the Lagrangian of
the action (\ref{eq:accion1}) takes the form
\begin{eqnarray}
 L=\pm
\sqrt{ \left(b \cdot \dot X\right)^{2}-\left(b\cdot b\right) \left(\dot X \cdot \dot X \right)}.
 \label{eq:accion3}
\end{eqnarray}
Thus, to restore the Lorentz symmetry we regard now the constant
vector $b^{M}$ as a local field $B^{M}=B^{M}(X),$ transforming under
Lorentz transformations as proper vector field, in this case the
action becomes
\begin{eqnarray}
S=\pm \int d\tau  \sqrt{ \left(B \cdot \dot X \right)^{2}-\left(B \cdot B\right)
 \left(\dot X \cdot \dot X\right)},
\end{eqnarray}
and this is Lorentz invariant.\\

Now this system has the primary constraints
\begin{eqnarray}
\Phi_{1}&=&P_{M}B^{M}=0, \\
\Phi_{2}&=&P_{M}B^{M} +B_{M}B^{M}=0.
\end{eqnarray}
In consequence the total Hamiltonian is
\begin{eqnarray}
H_{T}=\lambda_{1}\Phi_{1}+ \lambda_{2}\Phi_{2}.
\end{eqnarray}
Note that
\begin{eqnarray}
\{P_{M}B^{M}, P_{N}B^{N} +B_{N}B^{N}\}=2\left(P_{N}P_{M}-B_{N}B_{M}\right)\partial^{M}B^{L} .
\end{eqnarray}
From which it immediately follows that we will get more constraints and these depend on the form
of $B^{M}.$\\
An interesting case corresponds to assume that $B^{M}=X^{M},$ then
the action will be
\begin{eqnarray}
S=\pm \int d\tau  \sqrt{ \left(X \cdot \dot X \right)^{2}-\left(X \cdot X\right) \left(\dot X \cdot \dot X\right)},
\label{eq:accion2}
\end{eqnarray}
with primary constraints
\begin{eqnarray}
\Phi_{1}&=&P_{M}X^{M}=0, \label{eq:cons1}\\
\Phi_{2}&=&P_{M}X^{M} +X_{M}X^{M}=0, \label{eq:cons2}
\end{eqnarray}
furthermore
\begin{eqnarray}
\{\Phi_{1}, \Phi_{2}\}=2\Phi_{3}\qquad \Phi_{3}=\left(P_{M}P^{M} -X_{M}X^{M}\right).
\end{eqnarray}
Then, using the Dirac's method, we must satisfy that
\begin{eqnarray}
\Phi_{3}=\left(P_{M}P^{M} -X_{M}X^{M}\right)=0.
\end{eqnarray}
Now, these constraints satisfy the algebra
\begin{eqnarray}
\{\Phi_{1}, \Phi_{2}\}=2\Phi_{3},\quad \{\Phi_{1}, \Phi_{3}\}=2\Phi_{2} \quad
\{\Phi_{2}, \Phi_{3}\}=8\Phi_{1},
\end{eqnarray}
then, it follows that are first class constraints and there are no
more constraints. Thus it appears that the extended Hamiltonian is
\begin{eqnarray}
H_{E}&=&\lambda_{1} \left(P_{M}P^{M} +X_{M}X^{M}\right)  +\lambda_{2}  P_{M}X^{M}+
\lambda_{3} \left(P_{M}P^{M} -X_{M}X^{M}\right).
\end{eqnarray}
On the other hand, by defining
\begin{eqnarray}
\phi_{1} &=&\frac{1}{2}  P_{M}P^{M},\qquad  \phi_{2}=  P_{M}X^{M},\qquad
\phi_{3}=\frac{1}{2} X_{M}X^{M},\\
\gamma_{1}&=& \frac{ \lambda_{1}+\lambda_{2}}{2},\quad \gamma_{2}=\lambda_{2},\quad \gamma_{1}= \frac{ \lambda_{1}-\lambda_{2}}{2},
\end{eqnarray}
we obtain
\begin{eqnarray}
H_{E}= \gamma_{1} \phi_{1}+ \gamma_{2}  \phi_{2}+\gamma_{3}\phi_{3}.
\end{eqnarray}
This is the Hamiltonian of the two time physics
\cite{bars:gnus,me:gnus}. We know that the Lagrangian of the two
time physics has as a local symmetry the group $Sp(2)$ and as global
symmetry the conformal group \cite{me:gnus}. Then, the Hamiltonian
action of this system has more symmetries that the original action (\ref{eq:accion}).\\

Let us observe, finally, that the two time physics contains
different systems when we use only a temporal coordinate and acts
like a model that unifies the dynamics of different systems
\cite{bars:gnus}. In particular, it contains the relativistic free
particle. In consequence, by imposing the Lorentz invariance to the
Lagrangian (\ref{eq:accion3}) we obtain a system that unifies
different models to the level of point particle \cite{bars:gnus}.\\

Finally, it should be emphasized that by consistency, this system
requires two temporal coordinates and the signature must be of the
form ${\rm sig}(\eta)=(-,-,+,\cdots,+).$ Thus, to make sense of this
system is required that the signature had a transition from
 ${\rm sig}(\eta)=(-,+,+,\cdots,+)$ to ${\rm
sig}(\eta)=(-,-,+,\cdots,+).$  It is interesting to mention that
recently was proposed materials with this kind of characteristics \cite{2tmateriales:gnus}.\\

\section{Poincar\'e symmetry and relativistic string}

The action (\ref{eq:accion2}) is Lorentz invariant, but not
invariant under the Poincar\'e group. Now, to write down an explicit
Poincar\'e action we must choose $B^{M}$  in such way that it be
invariant under translations. Note that, if we take
$B^{M}=\frac{\partial C^{M}}{\partial \tau},$ the Lagrangian
(\ref{eq:accion3}) is invariant under Poincar\'e. However, not all
the equations of motion will be independent and corresponding system
must have constraints. In this case the Hamiltonian analysis is
quite involved. Another case, corresponds to take $C^{M}=X^{M},$
here the Lagrangian  (\ref{eq:accion3}) is invariant
under Poincar\'e, but vanishes.\\

Another way to recover the Poincar\'e invariance is to consider that
the action (\ref{eq:accion}) was built taking as starting point a
dispersion relation obtained in Field Theory, i.e., the action was
established from a simplification of Field theory to a point
particle. It must be stressed that the action for the two time
physics was obtained in the same way \cite{marnelius:gnus}. Then, to
recover de Poincar\'e invariance we will take the inverse path,
i.e., we will transform our particle model into a Field Theory. In
fact, assuming that the coordinates depend on an extra parameter
$\sigma,$ i.e. $X^{M}= X^{M}(\tau,\sigma),$ it is equivalent to
suppose that the particles are not
  points and instead are linear extended objects. In that case we can take $B^{M}= T \frac{
\partial X^{M}}{\partial \sigma},$ where $T$ is a constant, and the Lagrangian
(\ref{eq:accion3}) will be invariant under  Poincar\'e transformations.\\

We can use the expression $B^{M}= T \frac{ \partial X^{M}}{\partial
\sigma},$ in the constraints (\ref{eq:cons1})-(\ref{eq:cons2}) and
results
\begin{eqnarray}
\phi_{1}&=&P_{M} \frac{ \partial X^{M}}{\partial \sigma} =0, \label{p1}\\
\phi_{2}&=&\left(P_{M}\right)^{2}+T^{2} \frac{ \partial X^{M}}{\partial \sigma}\frac{ \partial X_{M}}{\partial \sigma} =0. \label{p2}
\end{eqnarray}
The equations (\ref{p1}) and (\ref{p2})  are the constraints of the
Hamiltonian action of the relativistic string \cite{cuerdas:gnus}.
For instance, using $B^{M}= T \frac{ \partial X^{M}}{\partial
\sigma}$ in the Lagrangian (\ref{eq:accion3}) we get
\begin{eqnarray}
L=T  \sqrt{ \left(\frac{ \partial X^{M}}{\partial \sigma} \frac{ \partial X_{M}}{\partial \tau}\right)^{2}- \left(\frac{\partial X^{N}}{\partial \sigma} \frac{ \partial X_{N}}{\partial \sigma} \right)
\left( \frac{\partial X^{M}} {\partial \tau} \frac{ \partial X_{M}}{\partial \tau} \right)}.
\end{eqnarray}
From this expression, the action takes the form
\begin{eqnarray}
S=T \int d\tau d\sigma  \sqrt{ \left(\frac{ \partial X^{M}}{\partial \sigma} \frac{ \partial X_{M}}{\partial \tau}\right)^{2}- \left(\frac{\partial X^{N}}{\partial \sigma} \frac{ \partial X_{N}}{\partial \sigma} \right)
\left( \frac{\partial X^{M}} {\partial \tau} \frac{ \partial X_{M}}{\partial \tau} \right)}.
\label{eq:nambu}
\end{eqnarray}
and this is the action of Nambu-Goto relativistic string \cite{cuerdas:gnus}.\\

In this way, by imposing the Lorentz and the Poincar\'e symmetries
to the Lagrangian (\ref{eq:accion3}) we get the action of a bosonic
string. It should be mentioned that to make Poincar\'e invariant the
Snyder space \cite{snyder:gnus}, Yang proposed an extra dimension
\cite{yang:gnus}. In our case we use a similar process, introducing
the coordinate $\sigma$ and we reestablish the Lorentz and
Poincar\'e symmetries. \\

Furthermore, using the Nambu-Goto (\ref{eq:nambu}) we can also recover the action (\ref{eq:accion2}).
In fact, in order to put this back, we shall use the expression
\begin{eqnarray}
\frac{ \partial X^{M}}{\partial \sigma}=\alpha X^{M}
\end{eqnarray}
then
\begin{eqnarray}
X^{M}(\tau,\sigma)= e^{\alpha \sigma} u^{M}(\tau).
\end{eqnarray}
Using this result in (\ref{eq:nambu}), we get
\begin{eqnarray}
S=\beta  \int d\tau  \sqrt{ \left(u\cdot \dot u\right)^{2}-\left(u \cdot u\right) \left(\dot u \cdot \dot u\right)},
\end{eqnarray}
with $\beta=T|\alpha|\int d\sigma e^{2\alpha \sigma}.$ This action
is equivalent to (\ref{eq:accion2}).

\section{Summary}
\label{s:Summ} In this work was analyzed several properties of the
particle with Lorentz symmetry violation recently proposed by
Kosteleck\'y. We introduced an alternative action for this system,
that can be interpreted as a particle in a curved space, where the
metric depends on the Lagrange multipliers. In addition, was shown
that there exist a limit where this system has more local symmetries
that usual relativistic particle. In this limit we saw that there
were several forms to reestablish the Lorentz symmetry. In
particular, for one of this forms we obtain a relationship with the
two time physics. Finally, by recovering the Poincar\'e symmetry the
action of the relativist string was obtained.

\end{document}